\documentclass[aps,prd,nofootinbib,longbibliography,reprint]{revtex4-2}
\usepackage{latexsym,bm,graphicx,color,xcolor,nicefrac,titletoc,dirtytalk,enumerate,mathtools,xfrac,xcolor,tensind,physics,amsmath,amssymb,mathtools}

\usepackage{mlmodern}
\usepackage[T1]{fontenc}

\usepackage[hidelinks]{hyperref}

\renewcommand{\thesection}{\arabic{section}}
\renewcommand{\thesubsection}{\arabic{section}.\arabic{subsection}}
\renewcommand{\thesubsubsection}{\arabic{section}.\arabic{subsection}.\arabic{subsubsection}}

\usepackage{titlesec}
\titleformat{\section}
{\centering\bfseries}{\thesection}{0.3em}{}
\titlespacing*{\section}{0pt}{2ex}{1ex}

\titleformat{\subsection}
{\normalsize\bfseries}{\thesubsection}{0.5em}{}
\titlespacing*{\subsection}{0pt}{1.2ex}{0.6ex}

\titleformat{\subsubsection}[runin]
{\normalsize}{\thesubsubsection}{0.3em}{}[ --]
\titlespacing*{\subsubsection}{0pt}{0.8ex}{0.5em}

\numberwithin{equation}{section}

\def\a{\alpha}

\def\g{\gamma}

\def\d{\delta}
\def\D{\Delta}

\def\ve{\varepsilon}
\def\f{\phi}

\def\l{\lambda}
\def\L{\Lambda}
\def\m{\mu}
\def\n{\nu}
\def\r{\rho}

\def\S{\Sigma}

\def\th{\theta}

\def\O{\Omega}

\def\pd{\partial}
\def\nab{\nabla}
\def\pr{\prime}

\newcommand{\sq}{\sqrt}
\newcommand{\sqdet}{\sq{-g}}

\newcommand{\rp}{r_+}

\newcommand{\be}{\begin{eqnarray}}
\newcommand{\ee}{\end{eqnarray}}

\newcommand{\cL}{\mathcal{L}}
\newcommand{\cM}{\mathcal{M}}

\newcommand{\tT}{\widetilde{T}}

\newcommand{\Qin}{Q_\text{in}}

\begin{document} 
	
\title{The Smarr Formula is Gauss's Law: A Kerr-Schild Single-Copy Perspective}
\author{G\"{o}khan Alka\c{c}}
\email{alkac@mail.com}
\affiliation{Department of Aerospace Engineering, Faculty of Engineering,\\ At{\i}l{\i}m University, 06836 Ankara, T\"{u}rkiye}

\begin{abstract}
In the Kerr-Schild double copy, static and spherically symmetric black hole solutions of general relativity are mapped to purely electric solutions of Maxwell's theory in flat spacetime. We demonstrate that, for these configurations, the thermodynamic Smarr formula is structurally identical to the single-copy Gauss's law. Extending this to asymptotically anti-de Sitter spacetimes, we prove that  the thermodynamic pressure-volume term naturally emerges from a gauge-theoretic background subtraction. This relationship establishes a novel connection between the classical double copy and black hole thermodynamics.
 \end{abstract}
\maketitle


\section{Introduction}
In light of the findings on perturbative scattering amplitudes, revealing a $\text{Gravity} = \text{Yang-Mills}^2$ type relationship \cite{Bern:2008qj,Bern:2010yg,Bern:2010ue}, it is unsurprising that perturbative solutions of gravity and Yang-Mills theories are related \cite{Luna:2016hge}. This inspired the study of exact solutions, and remarkably, it has been shown that one can map certain exact solutions of General Relativity (GR) to solutions of Maxwell's theory \cite{Monteiro:2014cda,CarrilloGonzalez:2017iyj,Alkac:2021bav,Luna:2018dpt,Godazgar:2020zbv,Easson:2021asd,Easson:2022zoh,Alkac:2023glx}. This correspondence has been extensively investigated under the name of the classical double copy, where the associated gauge theory solution is referred to as the single copy.

As the linearity of Maxwell's equations suggests, such a map requires the linearization of the Einstein equations for the gravitational solution under consideration. There are two main ways to achieve this\footnote{See \cite{Alkac:2024pfd,Frolov:2025ddw,Gumus:2026dpv} for a recent mini-superspace formulation, which is still in its infancy.}. In the metric formulation, known as the Kerr-Schild  double copy \cite{Monteiro:2014cda,CarrilloGonzalez:2017iyj,Alkac:2021bav}, one takes advantage of the fact that the trace-reversed Einstein equations are linear in the perturbation for spacetimes admitting Kerr-Schild coordinates \cite{Gurses:1975vu}. In the spinor formulation, the Weyl spinor of the gravitational solution, which has a linear character in the perturbation, is related to the field strength spinor of a solution of Maxwell's theory defined on a suitable background spacetime \cite{Luna:2018dpt,Godazgar:2020zbv,Easson:2021asd,Easson:2022zoh,Alkac:2023glx}. When both are applicable to a given solution, they have been shown to be consistent, even for certain exotic cases \cite{Alkac:2023glx,Alkac:2025iyw}.

Fortunately, the classical double copy is applicable to many physically relevant solutions of GR, demonstrating that it is not merely an interesting mathematical correspondence but possesses the potential to provide physical insight. One of the most important classes of solutions is undoubtedly black holes. While their metrics can generally be expressed in Kerr-Schild coordinates to obtain the corresponding gauge theory solutions, a direct mapping between the physical properties of the two systems has remained elusive. For example, key thermodynamic properties of a black hole, such as its temperature and entropy, are fundamentally defined by the presence of an event horizon. Conversely, on the gauge theory side, this surface is simply a submanifold in the flat background with no inherent physical significance. This disparity raises a natural question: how, then, can the intrinsic thermodynamic features of a black hole be reflected in its single-copy gauge field?

Here, we tackle this problem by revisiting a result of central importance in black hole physics: the Smarr formula \cite{Smarr:1972kt}. As famously stated by Wheeler, a black hole has ``no hair'' \cite{Ruffini:1971bza}; that is, it is uniquely characterized by its global charges, such as mass, angular momentum, and electric charge. The Smarr formula relates these elegantly in a form where the total energy, the mass, is expressed as the sum of a ``surface energy'' and other forms of energy. While the surface energy is proportional to the product of the temperature and the entropy, the remaining terms are typically formed by the products of global charges (other than the mass) and their corresponding conjugate potentials. This formula is a cornerstone of gravitational thermodynamics; however, its potential manifestation on the gauge theory side has remained unexplored. Although previous studies have observed a local correspondence between the gravitational matter distribution and the electric charge density for horizonless Kerr-Schild geometries \cite{Ridgway:2015fdl}, the implications for global surface integrals and the resulting black hole thermodynamics have not been addressed.

In this work, we bridge this conceptual gap by demonstrating that the surface energy of a black hole has a precise counterpart in its single copy. To establish this framework, we focus strictly on static, spherically symmetric black hole solutions. Because they lack the complications arising due to rotation, these spacetimes provide the cleanest environment to expose the fundamental structural equivalence between the black hole Smarr formula and single-copy Gauss's law. Specifically, we show that the surface energy, which is originally a purely gravitational quantity associated with the event horizon, is directly proportional to the electric flux of the gauge field calculated over the corresponding surface in the flat background spacetime. This relationship allows us to interpret the black hole's thermodynamic Smarr formula as a gauge-theoretic identity. The ``missing'' physical information is encoded in a set of flux-charge relations; while the total energy is captured by the electric flux at infinity, the remaining terms in the Smarr formula are identified with the charge distributions integrated over the spatial region extending from the surface corresponding to the event horizon to infinity. This result provides a natural complement to recent developments in the quantum regime, where it has been shown that Hawking radiation and its associated thermality emerge from the double copy of particle production in background gauge fields \cite{Ilderton:2025aql, Aoude:2025jvt} (see also \cite{Carrasco:2025bgu}). 

The remainder of this paper is organized as follows: In Sec.~\ref{sec:KS}, we briefly review the essential features of the Kerr-Schild double copy. We then use the Schwarzschild solution to demonstrate the fundamental mapping between the Smarr formula and Gauss's law in Sec.~\ref{sec:schw}, before extending this analysis to the Reissner-Nordstr\"om black hole in Sec.~\ref{sec:RN}. The inclusion of a cosmological constant is discussed in Sec.~\ref{sec:cosm}. In Sec.~\ref{sec:bardeen}, we apply our framework to regular black holes by analyzing the Bardeen spacetime with non-linear matter sources. Finally, we offer concluding remarks and discuss future directions in Sec. \ref{sec:conc}.

\section{The Kerr-Schild Double Copy}\label{sec:KS}
We consider a general action of the form
\begin{equation}\label{act}
I = I_g + I_m,
\end{equation}
where the gravitational and matter parts respectively read
\begin{equation}
\begin{aligned}
I_g &= \frac{1}{16 \pi} \int \dd^4 x \sqdet \left(R - 2\L\right) \\
I_m &= \frac{1}{16 \pi} \int \dd^4 x\sqdet\, \cL_m.
\end{aligned}
\end{equation}
We set $G=1$ for simplicity. $\L$ is the cosmological constant and $\cL_m$ is the matter Lagrangian. Einstein equations for the metric following from the action \eqref{act} are
\begin{equation}\label{eins}
G_{\m\n} + \L g_{\m\n} =8\pi T_{\m\n},
\end{equation}
where the energy-momentum tensor $T_{\m\n}$ is defined as
\begin{equation}
T_{\m\n} = -\frac{2}{\sqdet} \fdv{I_m}{g^{\m\n}}.
\end{equation}
If a metric can be written in the Kerr-Schild (KS) coordinates as follows
\begin{equation}\label{metKS}
g_{\m \n}=\eta_{\mu \nu}+\f\, k_\mu k_\nu, 
\end{equation}
where $\eta_{\m\n}$ is the flat background metric, $\f$ is the KS scalar and the vector $k_\m$ is null and geodesic with respect to the full and the background metric, Einstein equations \eqref{eins} are linear in the perturbation \cite{Gurses:1975vu}. A direct calculation of the Ricci tensor with mixed indices for a metric in \eqref{metKS} yields \cite{Stephani:2003tm}
\begin{equation}\label{Ricci}
R^\m_{\ \n}=\frac{1}{2} \Big[\partial^\a \partial^\m\left(\f\, k_\n k_\a\right)+\partial^\a \partial_\n\left(\f\, k^\m k_\a\right)
- \partial^2 \left(\f\, k^\m k_\n\right)\Big].
\end{equation}
Using this in the trace-reversed Einstein equations, which are given by
\begin{equation}\label{reversed}
R^\m_{\ \n} = 8 \pi \left[\tT^\m_{\ \n} + \frac{\L}{8\pi} \d^\m_{\ \n} \right],
\end{equation}
where
\begin{equation}
\tT^\m_{\ \n} = T^\m_{\ \n} - \frac{1}{2} \d^\m_{\ \n} T^\a_{\ \a},
\end{equation}
one obtains a linear equation for the KS scalar $\f$, which is the only unknown in the metric in \eqref{metKS}.

For a stationary metric ($\partial_0 g_{\mu \nu}=0$), by identifying $\f k_\m \equiv A_\m$ and choosing $k_0 = +1$, one finds
\begin{equation}
R^\m_{\ 0} = - \frac{1}{2} \pd_\a F^{\a\m}, \quad F_{\m\n} = 2\, \pd_{[\m} A_{\n]}.
\end{equation}
Therefore, from the $\m0$-components of the trace-reversed Einstein equations \eqref{reversed}, we obtain the Maxwell equations for the single copy field $A_{\m}$ defined on a flat spacetime background \cite{Monteiro:2014cda}
\begin{equation}\label{max}
\pd_\a F^{\a \m} = 4 \pi J^\m,
\end{equation}
where the gauge theory source $J^\m$ is related to the trace-reversed energy momentum tensor via
\begin{equation}\label{JtoT}
J^\m = -4\, \left[\tT^\m_{\ 0}+\frac{\L}{8 \pi} \d^\m_{\ 0}\right].
\end{equation}
The time component of \eqref{max} reduces to the Poisson equation on flat spacetime
\begin{equation}
\nab^2 \f = - 4\pi \r,
\end{equation}
where $\r$ is the charge density.

We now proceed to apply this formalism to specific gravitational solutions. In the following section, we provide a general treatment of the Kerr-Schild double copy for static, spherically symmetric black hole metrics and analyze the Schwarzschild solution, establishing the equivalence of the Smarr formula to Gauss’s law.

\section{Schwarzschild Black Hole: Surface Energy and Gauss's Law}\label{sec:schw}
The line element of a general static, spherically symmetric black hole solution in the Boyer-Lindquist coordinates is given by
\begin{equation}\label{BL}
\dd{s}^2 = - f(r) \dd{t}^2+\frac{1}{f(r)} \dd{r}^2+ r^2 \dd{\O}^2,
\end{equation}
where $\dd{\O}^2$ is the metric of a $2$-dimensional unit sphere. Applying the transformation
\begin{equation}
\dd{t} \to \dd{t} - \frac{1-f(r)}{f(r)} \dd{r},
\end{equation}
one obtains the metric in the KS coordinates \eqref{metKS} with a flat background metric expressed in spherical coordinates. The KS scalar $\f$ and the vector $k_\m$ are given by
\begin{equation}
\f = 1 - f, \quad  k_\m\dd{x}^\m = \dd{t}+\dd{r}.
\end{equation}
For such a solution, the only non-vanishing components of the single-copy field strength tensor $F_{\mu\nu}$ correspond to a radial electric field
\begin{equation}\label{radE}
\vb{E} = E\, \vu{r}, \qquad E = F_{0r} = - \f^\pr,
\end{equation}
where the prime denotes differentiation with respect to $r$. Maxwell's equations \eqref{max} reduce to
\begin{equation}\label{Gdif}
\div{\vb{E}} = 4 \pi \r,
\end{equation}
which is just the differential form of Gauss's law. To relate this to the global charges of the black hole solution, we consider the integral form of Gauss's law, whose most familiar form reads
\begin{equation}\label{Gint}
\oint_{\pd\S} \vb{E} \vdot \dd{\vb{A}} = 4 \pi \Qin, \quad \Qin =\int_\S \r \dd{V},
\end{equation}
which states that the flux of the electric field $\vb{E}$ over the closed surface $\pd\S$ is proportional to the charge contained in the volume $\S$ enclosed by the surface $\pd\S$.

The Schwarzschild solution is characterized by the metric function
\begin{equation}
	f = 1-\frac{2m}{r},
\end{equation}
where $m$ denotes the mass. It is a vacuum solution ($\cL_m=0$) with no cosmological constant ($\L=0$). Applying the KS prescription, we find the single copy electric field and the corresponding charge density as 
\begin{equation}
	E = \frac{2m}{r^2}, \qquad \r = 2m \d(\vb{r}).
\end{equation}
This corresponds to the Coulomb solution for a point charge $2m$, directly satisfying Gauss's law in \eqref{Gint}.

As is well-known, the temperature $T$ and the entropy $S$ of a black hole are proportional to the surface gravity and the area of the event horizon respectively \cite{Hawking:1975vcx, Bekenstein:1973ur}. For a black hole solution with the line element \eqref{BL}, they read
\begin{equation}
	T = \frac{f^\pr(r_+)}{4\pi} = - \frac{\f^\pr(r_+)}{4\pi}= \frac{E(r_+)}{4\pi}, \quad S = \frac{A(\rp)}{4},
\end{equation}
where $\rp$ is the location of the outer event horizon and $A(\rp) = 4 \pi \rp^2$ is the area of the event horizon. They satisfy the first law of black hole thermodynamics \cite{Bardeen:1973gs}
\begin{equation}\label{1stlawsch}
	\d m = T \d S.
\end{equation}

The product $2TS$, which corresponds to the surface energy of the black hole, is determined by the value of the single-copy electric field $E(r_+)$ and the area $A(r_+)$. Consequently, the surface energy is directly proportional to the electric flux evaluated at the surface $r=r_+$. We emphasize that while this integration is performed over an ordinary two-sphere within the flat Minkowski background, the area remains $A(r_+) = 4\pi r_+^2$, as in the original curved spacetime. Using the Gauss's law in \eqref{Gint}, we can write 
\begin{equation}
		2 T S = \frac{1}{8 \pi} \oint_{\mathclap{\qquad r=\rp}}\,\,  \vb{E} \vdot \dd{\vb{A}}=\frac{1}{2} \Qin,
\end{equation}
which yields
\begin{equation}\label{smarrsch}
	2 T S = m.
\end{equation}
This is the Smarr formula for the Schwarzschild black hole. Its usual derivation follows from Euler's theorem for homogeneous functions \cite{Smarr:1972kt}. Observing that the mass is a homogeneous function of entropy with degree $1/2$, and using $T = \pdv{m}{S}$ from the first law \eqref{1stlawsch}, one obtains Eq. \eqref{smarrsch}. Here, we obtain it on the gauge theory side from the integral form of Gauss's law. While the Schwarzschild black hole might seem too simple to draw a general conclusion, we will demonstrate that this precise structural equivalence persists for more general spacetimes.

\section{Reissner-Nordstr\"om Black Hole: Divergences, Generalized Gauss's Law and Komar Integrals}\label{sec:RN}
To test the generality of our result for the Schwarzschild black hole, we now consider a solution possessing an additional global charge.  The Reissner-Nordstr\"om (RN) black hole solution is characterized by the metric function
\begin{equation}\label{fRN}
f = 1 -\frac{2m}{r} + \frac{q^2}{r^2},
\end{equation}
where $m$ and $q$ are the mass and the electric charge of the black hole respectively.  The solution arises from coupling general relativity with zero cosmological constant ($\L = 0$) to Maxwell's theory with the Lagrangian\footnote{Note that $a_\m$ and $f_{\m\n}$ are the gauge field and the field strength on the gravity side. The single copy gauge field and field strength are distinguished by capital letters, and denoted by $A_\m$ and $F_{\m\n}$.}
\begin{equation}
\cL_m = -\frac{1}{16 \pi} f_{\m\n} f^{\m\n}, \quad  f_{\m\n} = 2\, \pd_{[\m} a_{\n]},
\end{equation}
and the gauge field $a_\m$ for this solution takes the form
\begin{equation}
a_\m \dd{x}^\m = \frac{q}{r} \dd{t}.
\end{equation}
From the metric function \eqref{fRN}, the corresponding electric field and the charge density can be easily found as
\begin{equation}\label{ErhoRN}
E = \frac{2m}{r^2} - \frac{2 q^2}{r^3}, \quad \r = 2m\d(\vb{r}) + \frac{q^2}{2 \pi r^4},
\end{equation}
which satisfy the differential form of Gauss's law \eqref{Gdif}.

As we see, the ``Schwarzschild-term'' in the metric function \eqref{fRN} yields the Coulomb solution with electric charge $2m$. In contrast, the term proportional to $\flatfrac{1}{r^2}$, which arises from the electric charge $q$ of the black hole, is sourced by a non-trivial charge distribution. When we naively check the integral form of Gauss's law in \eqref{Gint} for the surface $r=\rp$, the flux term reads
\begin{equation}\label{fluxRN}
\oint_{\mathclap{\quad \,\,\,\,r=\rp}}\,\,  \vb{E} \vdot \dd{\vb{A}} = 4 \pi \left(2m - \frac{2q^2}{\rp}\right).
\end{equation}
However, the total charge inside the surface $r=\rp$ diverges as follows
\begin{equation}\label{QinRN}
\Qin = 4 \pi \int_\ve^{\rp} \r\, r^2 \dd{r} = 2m - \frac{2 q^2}{\rp} +  \frac{2 q^2}{\ve} \text{ as } \ve \to 0.
\end{equation}
Evidently, the integral form of Gauss's law in \eqref{Gint} does not hold. However, this discrepancy does not present a formal issue, as the integral form remains valid provided the integrals on both sides are well-defined.

For a general form of Gauss's law applicable to more general charge distributions, let us consider the integral of the source vector $J^\m$ over the spatial volume $\S_\m$ and apply Stokes' theorem
\begin{equation}\label{gaussgen}
	\int_\S  J^\m \dd \S_\m = \frac{1}{4\pi} \int_\S  \pd_\n F^{\n\m} \dd \S_\m = \frac{1}{8 \pi} \oint_{\pd\S} F^{\m\n} \dd \S_{\m\n} ,
\end{equation}
where ${\pd\S}$ is the surface enclosing the volume $\S$. The volume and the surface elements are given respectively as $\dd \S_\m = n_\m \dd V = n_\m \sqrt{h}\, \dd x^3$ and  $\dd \S_{\m\n} = 2 n_{[\m}r_{\n]}  \dd S =2 n_{[\m}r_{\n]} \sqrt{\g}\, \dd x^2$. $n_\m = \pd_\m t$ and $r_\m = \pd_\m r$ are unit normal vectors, and $h$ and $\g$ are the determinant of induced metrics. We work in a spacetime covariant form since this will turn out to be convenient later. Choosing $\S$ to be the spatial region between the surfaces $r=\rp$ and $r \to \infty$, we find
\begin{align}
	\frac{1}{8 \pi} \oint_{\mathclap{\qquad r=\rp}}\,\, F^{\m\n} \dd \S_{\m\n} &= \frac{1}{8 \pi} \oint_{\mathclap{\qquad r \to \infty}}\,\, F^{\m\n} \dd \S_{\m\n} - 	\int_\S  J^\m \dd \S_\m,\label{FStokes}\\
	 \frac{1}{4 \pi} \oint_{\mathclap{\qquad r=\rp}}\,\, \vb{E} \vdot \dd{\vb{A}} &= \frac{1}{4 \pi} \oint_{\mathclap{\qquad r \to \infty}}\,\, \vb{E} \vdot \dd{\vb{A}} - \int_\S  \r\, \dd V.
\end{align}
Now, the charge integral computes the total charge between the surfaces $r = \rp$ and $r \to \infty$ and there is no divergence. The Dirac delta source does not contribute to the integral and the effect of the Schwarzschild/Coulomb term is accounted for via the flux at infinity. Calculating the surface energy from the flux at $r=\rp$ gives
\begin{align}
		2TS&= \frac{1}{8 \pi} \oint_{\mathclap{\qquad r=\rp}}\,\, \vb{E} \vdot \dd{\vb{A}}=  \frac{1}{8 \pi} \oint_{\mathclap{\qquad r \to \infty}}\,\, \vb{E} \vdot \dd{\vb{A}} -\frac{1}{2} \int_\S  \r\, \dd V,\label{TStoflux}\\
		&= m- \frac{q^2}{\rp}.
\end{align}
This is the Smarr formula for the RN black hole \cite{Smarr:1972kt}
\begin{equation}\label{SmarrRN}
	m= 2 T S + \Phi q,
\end{equation}
where 
\begin{equation}
	\Phi = \pdv{m}{q} = \frac{q}{\rp},
\end{equation}
is the chemical potential corresponding to the electric charge $q$, and the first law is now \cite{Bardeen:1973gs}
\begin{equation}
	\d m = T \d S + \Phi \d q.
\end{equation}
The Smarr formula \eqref{SmarrRN} can again be obtained from Euler's theorem of homogeneous functions by taking into account that the degree of $q$ is $1$. However, here, it follows from the calculation of the electric flux at the surface $r=\rp$ as long as the integral form of Gauss's law is employed appropriately.

To understand why this derivation works, let us recall the derivation of Smarr formula from Komar integrals \cite{Komar:1958wp,Wald:1984rg,Townsend:1997ku,Poisson:2009pwt}. For any Killing vector $\xi_\m$, we can define a Komar two-form $K_{\m\n} = 2\nab_{[\m} \xi_{\n]}$. Using the identity $\nab_\n K^{\n\m} = -2 R^\m_{\ \n} \xi^\n = 16 \pi \tT^\m_{\ \n} \xi^\n$ (when $\L=0$), we can write
\begin{equation}\label{KStokes}
	\frac{1}{4 \pi} \oint_{\mathclap{\qquad r=\rp}}\,\, K^{\m\n} \dd \S_{\m\n} = \frac{1}{4 \pi} \oint_{\mathclap{\qquad r \to \infty}}\,\, K^{\m\n} \dd \S_{\m\n} + 4	\int_\S  \tT^\m_{\ \n} \xi^\n \dd \S_\m,
\end{equation}
where we have used Stokes' theorem. Note that this is now a calculation in curved spacetime and we work in the Boyer-Lindquist coordinates. Unit normals are $n_\m = \sqrt{f}  \pd_\m t$ and $r_\m = \frac{1}{\sqrt{f}} \pd_\m r$. Explicit calculation of each term in this equation by choosing the time-like Killing vector $\xi_{(t)} = \pd_t$ yields the Smarr formula for a given solution. For the RN black hole, it yields
\begin{equation}
	4 T S = 2m - \frac{2q^2}{\rp}.
\end{equation}
This also leads to the definition of the Komar mass as 
\begin{equation}\label{KomarMass}
	\cM = \frac{1}{8 \pi}  \oint_{\mathclap{\qquad r \to \infty}}\quad K^{\m\n}_{(t)} \dd \S_{\m\n} = m,
\end{equation}
 where $K^{(t)} _{\m\n} = 2\nab_{[\m} \xi^{(t)}_{\n]}$ is the Komar two-form corresponding to the time-like Killing vector $\xi_{(t)} = \pd_t$. Here, we have chosen the overall constant such that it is consistent with the single-copy calculation.

Eq. \eqref{KStokes} with the time-like Killing vector $\xi_{(t)} = \pd_t$ and Eq. \eqref{FStokes} are indeed equivalent to each other not only for the RN black hole but more generally. First of all, the volume element $\dd \S_\m$ is the same in curved spacetime and in flat background spacetime because  $\dd \S_{\text{(curved)}} = \frac{1}{\sqrt{f}} \dd \S_{\text{(flat)}}$. As a result, the volume integrals in both equations are identical since $J^\m = -4 \tT^{\m}_{\ 0}$. The surface elements $\dd \S_{\m\n}$ also take the same form since $\dd S_{\text{(curved)}} = \dd S_{\text{(flat)}}$ and $\sqrt{f}$ terms in unit normals $n_\m$ and $r_\m$ cancel each other. For the Komar two-form $K_{\m\n}$, let us first find the covariant components of the time-like Killing vector $\xi_{(t)} = \pd_t$ as follows
\begin{equation}
	\xi_\m^{(t)} = g_{\m\n} \xi^\n_{(t)} = \left(-1+\f,0,0,0\right),
\end{equation}
where we have used $f = 1-\f$. It differs from the single copy gauge field $A_\m = (\f,\f,0,0)$ only by a pure gauge [$A_\m = \xi_\m^{(t)} + \pd_\m \l, \l = t+\int \f(r) \dd{r}$]. Therefore, their field strength tensors are equal, which implies
\begin{equation}
	K_{\m\n}^{(t)} =  F_{\m\n},
\end{equation}
since $K^{(t)}_{\m\n} = 2 \nab_{[\m} \xi^{(t)}_{\n]} = 2 \pd_{[\m} \xi^{(t)}_{\n]}$. This equality also holds for the contravariant components because $g^{00} g^{11} =\eta^{00} \eta^{11}= -1$ for metrics that we study. As a result, we have also shown that the surface integrals in Eqs. \eqref{FStokes} and \eqref{KStokes} are also equal, completing the proof of the full equivalence.

We note that the equivalence between the single-copy gauge field and spacetime Killing vectors, leading to the relation $K_{\m\n}^{(t)} = F_{\m\n}$, was observed in \cite{Caceres:2025eky} while investigating the geometric unification of Kerr-Schild and Weyl double copies and gravito-electromagnetism. However, the physical focus in that work is restricted to asymptotic charges and tidal tensors. Here, we uncover the thermodynamic significance of this identity: by applying Stokes' theorem across the exterior spatial domain bounded by the event horizon and spatial infinity, the geometric Komar identity directly translates into single-copy Gauss's law, with the horizon flux encapsulating the black hole surface energy $2TS$.
\section{Cosmological Constant and Black Hole Enthalpy}\label{sec:cosm}
We now extend our analysis to include a negative cosmological constant $\L = -\flatfrac{3}{\ell^2}$, where $\ell$ is the anti-de Sitter (AdS) radius. Following our analysis of the RN black hole, the metric function is modified as
\begin{equation}
f = 1-\frac{2m}{r}+\frac{q^2}{r^2}+\frac{r^2}{\ell^2},
\end{equation}
giving rise to the following electric field and charge density
\begin{equation}
\begin{aligned}
E &= \frac{2m}{r^2}-\frac{2q^2}{r^3}+ \frac{2 r}{\ell^2},\\
\r &= 2m \d(\vb{r})+\frac{q^2}{2\pi r^4}+\frac{3}{2 \pi \ell^2}.
\end{aligned}
\end{equation}
We see that the effect of a negative cosmological constant is an electric field linearly increasing with the radial coordinate $r$ sourced by a constant charge density filling all space.

Let us calculate the surface energy by using the prescription in \eqref{TStoflux}. The terms at the right-hand side are found as
\begin{equation}
	 \begin{aligned}
	  \frac{1}{8 \pi} \oint_{\mathclap{\qquad r \to \infty}}\,\, \vb{E} \vdot \dd{\vb{A}} &= m + \frac{R^3}{\ell^2} \text{ as } R \to \infty,  \\
	  \frac{1}{2} \int_\S  \r\, \dd V &= \frac{q^2}{\rp}+\frac{R^3-\rp^3}{\ell^2} \text{ as } R \to \infty.
	 \end{aligned}
\end{equation}
Both are divergent, but in an identical fashion. Hence, their difference, the surface energy, is finite and reads
\begin{equation}
	2 T S = m-\frac{q^2}{\rp}+\frac{\rp^3}{\ell^2}.
\end{equation}

It is customary to define the thermodynamic pressure as
\begin{equation}\label{Pdef}
P = -\frac{\L}{8 \pi},
\end{equation}
and write the Smarr relation as
\begin{equation}\label{Smarrmod}
m = 2 T S + \Phi q -2 P V,
\end{equation}
where $V$ is the conjugate thermodynamic volume defined by
\begin{equation}\label{Vdef}
V = \pdv{m}{P} = \frac{4 \pi \rp^3}{3}.
\end{equation}
Motivated by this, one can extend the first law with a variable cosmological constant as follows
\begin{equation}
\d m = T \d S + \Phi \d q +V\d P,
\end{equation}
which suggests that the mass $m$ can now be interpreted as the enthalpy of the black hole \cite{Kastor:2009wy} (See \cite{Dolan:2010ha,Cvetic:2010jb} for other pioneering works). From these identifications, entirely new phase transitions emerge, resulting in the so-called ``black hole chemistry''. See \cite{Mann:2025xrb} for a recent review.

Remarkably, these divergences do not preclude the derivation of the Smarr formula. However, a naive cancellation of the divergences is somewhat misleading. To establish a rigorous one-to-one identification of the thermodynamic variables, the mass $m$ must emerge exclusively from the flux integral at spatial infinity, while the other forms of energy should follow from the charge integral in the bulk. 

In the standard Komar formalism, the inclusion of the cosmological constant is accounted for by the shift $\tT^\m_{\ \n} \to \tT^\m_{\ \n} + \flatfrac{\L}{8 \pi} \d^\m_{\ \n}$. Due to this background contribution, the standard mass defined in \eqref{KomarMass} becomes ill-defined because of the non-asymptotic flatness of the spacetime. One solves this issue by modifying the Komar two-form as $K_{\m\n} \to K_{\m\n} + 2 \L w_{\m\n}$, where $w_{\m\n}$ is the Killing potential satisfying $\xi^\mu = \nabla_\nu w^{\n\m}$ \cite{Bazanski:1990qd,Kastor:2008xb}. 

In our gauge theory formalism, this geometric modification corresponds to subtracting the cosmological constant contributions from both the flux term and the charge integral. Effectively, this is a background subtraction. The cosmological constant manifests as a constant background charge density $\r_{\text{bg}}$ sourcing a background electric field $E_{\text{bg}}$. Applying Gauss's law to the regularized field $E_{\text{reg}} = E - E_{\text{bg}}$ and the regularized source $\r_{\text{reg}}  = \r - \r_{\text{bg}}$ over the volume $\S$ (from $r=\rp$ to $r \to \infty$), we can express the flux at the horizon as
\begin{equation}\label{regGauss}\frac{1}{8 \pi} \oint_{\mathclap{\qquad \,\, r=\rp}}\,\,  \vb{E_{\text{reg}}} \vdot \dd{\vb{A}} = \frac{1}{8 \pi} \oint_{\mathclap{\qquad \,\, r \to \infty}}\,\,  \vb{E_{\text{reg}} } \vdot \dd{\vb{A}} - \frac{1}{2} \int_\S\r_{\text{reg}}  \dd{V}.
\end{equation}
The terms on the right-hand side can now be identified precisely. At spatial infinity, the flux of the regularized field isolates the finite mass, directly yielding $m$. In the bulk integral over $\S$, the subtraction of the background charge $\r_{\text{bg}}$ ensures that the volume integral strictly isolates the chemical potential term, $\flatfrac{q^2}{\rp} = \Phi q$. On the left-hand side, the thermodynamic volume term emerges from the boundary condition at the horizon. The physical temperature $T$ of the AdS black hole is determined by the full, unregularized field $E$ at $\rp$, meaning the total surface energy is $2TS = \frac{1}{8\pi} \oint_{\rp} \mathbf{E} \cdot \dd\mathbf{A}$. Therefore, the flux of the \textit{regularized} field at the horizon is the total surface energy minus the background flux. Evaluating this background flux at $\rp$ yields exactly $\frac{1}{8\pi} \oint_{\rp} \mathbf{E}_{\text{bg}} \cdot \dd\mathbf{A} = \flatfrac{\rp^3}{\ell^2} = 2PV$. Consequently, the regularized flux at the horizon evaluates to $2TS - 2PV$. Equation \eqref{regGauss} thus yields
\begin{equation}
	2TS - 2PV = m - \Phi q,
\end{equation}
perfectly reproducing the rearranged Smarr formula \eqref{Smarrmod} and demonstrating a precise, term-by-term structural equivalence with the modified Komar prescription.

\section{Bardeen Black Hole: Non-Linear Matter Sources and Modified Smarr Relations}\label{sec:bardeen}
Having addressed the inclusion of a cosmological constant, we now test our setup against a more intricate matter sector: regular black holes. We consider the Bardeen black hole \cite{Bardeen:1968}, the first regular black hole solution proposed in the literature. Remarkably, this spacetime can be dynamically generated as an exact solution of Einstein gravity coupled to a specific model of non-linear electrodynamics described by the matter Lagrangian  \cite{AyonBeato:2000mjt}
\begin{equation}\label{LmBardeen}
	\cL_m = -\frac{3}{8sg^2} \left[\frac{\sqrt{ 2 g^2 f_{\m\n} f^{\m\n}}}{1+\sqrt{ 2 g^2 f_{\m\n} f^{\m\n}}} \right]^{5/2},
\end{equation}
where $s=\frac{|g|}{2m}$, while $g$ and $m$ are constants.

Assuming a static magnetic monopole configuration with magnetic charge $g$,
\begin{equation}
	a_\m \dd{x}^\m = -g \cos \th \dd{\f},
\end{equation}
the authors of \cite{AyonBeato:2000mjt} showed that the field equations yield the regular Bardeen metric function
\begin{equation}
	f = 1-\frac{2 m r^2}{\left(r^2+g^2\right)^{3 / 2}},
\end{equation}
where $m$ and $g$ represent the mass and the magnetic charge of the black hole, respectively.

Because the matter Lagrangian \eqref{LmBardeen} explicitly contains the global charges $m$ and $g$, the standard first law and Smarr formula receive non-trivial modifications. In \cite{Zhang:2016ilt}, this structure was systematically analyzed using the covariant Komar integral and its variation, leading to
\begin{equation}\label{D}
	\begin{aligned}
		\d m &= T \d S + \Psi \d g + K_g \d g + K_m \d m,\\
		m &= 2 T S + \D, \qquad \D = \Psi g + K_g g + K_m m,
	\end{aligned}
\end{equation}
where $\Psi$ denotes the magnetic potential conjugate to $g$, and $K_g, K_m$ represent the potentials conjugate to the parameters $g$ and $m$ appearing explicitly in the matter Lagrangian:
\begin{equation}
	\begin{aligned}
		\Psi &=\frac{3 m}{2 g}\left[1-\frac{\rp^5}{\left(\rp^2+g^2\right)^{5 / 2}}\right],\\
		K_g &= \frac{3 m}{2 g}\left[\frac{\rp^3(\rp^2+2 g^2)}{\left(\rp^2+g^2\right)^{5 / 2}}-1\right], \\
		K_m &= 1-\frac{\rp^3}{\left(\rp^2+g^2\right)^{3 / 2}}.
	\end{aligned}
\end{equation}
As we see, the first law and the Smarr formula now possess additional terms due to the existence of the parameters $m$ and $g$ in the Lagrangian \eqref{LmBardeen}.

Let us apply our formalism to check whether the $\D$-term, which accounts for the energy contributions other than the surface energy $2TS$, can be reproduced. Applying the Kerr-Schild prescription, we find the single-copy electric field and charge density as
\begin{equation}
	E = \frac{2 m r\left(r^2 -2 g^2\right)}{\left(r^2+g^2\right)^{5 / 2}}, \quad \r = \frac{3 m g^2 \left(3 r^2 -2 g^2\right)}{2 \pi\left(r^2+g^2\right)^{7 / 2}}.
\end{equation}
Using Eq. \eqref{TStoflux}, we obtain
\begin{equation}
	2TS= \frac{1}{8 \pi} \oint_{\mathclap{\qquad r \to \infty}}\,\, \vb{E} \vdot \dd{\vb{A}} -\frac{1}{2} \int_\S \r\, \dd V = m - \D,
\end{equation}
where
\begin{equation}\label{D2}
	\D = m \left[1-\frac{\rp^3\left(\rp^2-2 g^2\right)}{\left(\rp^2+g^2\right)^{5 / 2}}\right].
\end{equation}
The mass $m$ can be expressed in terms of the event horizon radius $\rp$ via the horizon condition $f(\rp) = 0$ as
\begin{equation}
	m = \frac{\left(\rp^2+g^2\right)^{3 / 2}}{2 \rp^2}.
\end{equation}
Using this relation, it is straightforward to show that the $\D$-terms in \eqref{D} and \eqref{D2} are identical. This demonstrates the robustness of our construction, even in highly non-trivial settings involving non-linear matter sources.

\section{Conclusion and Future Directions}\label{sec:conc}

In this work, we have demonstrated a direct, physical correspondence between the thermodynamic properties of black holes and the classical electrodynamics of their Kerr-Schild single copies. By focusing on static, spherically symmetric spacetimes, we have shown that the Smarr formula, a cornerstone of black hole thermodynamics, is not merely analogous to, but structurally identical to the generalized Gauss's law in the single-copy gauge theory. 

The most striking realization of this framework is the strict equality between the Komar two-form of the spacetime and the field strength tensor of the flat-space gauge theory ($K_{\m\n} = F_{\m\n}$). This identity allows us to map purely geometric surface integrals directly to electromagnetic fluxes. We summarize the exact dictionary established in this paper in Table \ref{tab:dictionary}.
\begin{table}[h]
	\caption{\label{tab:dictionary} The correspondence dictionary between black hole thermodynamics and single-copy electrodynamics.}
	\begin{ruledtabular}
		\small
		\setlength{\tabcolsep}{4pt}
		\begin{tabular}{p{4.2cm} p{4.2cm}}
			\textbf{Gravity} & \textbf{Gauge Theory} \\
			\colrule
			Event Horizon Surface in Curved Spacetime & Spherical Surface $r=r_+$ in Flat Spacetime \\
			Surface Energy ($2TS$) & Flux at $r_+$: $\frac{1}{8 \pi} \oint_{\mathclap{\qquad r =\rp}}\,\, \vb{E} \vdot \dd{\vb{A}}$ \\
			Mass / Enthalpy ($m$) & Asymptotic Flux:  $\frac{1}{8 \pi} \oint_{\mathclap{\qquad r \to \infty}}\,\, \vb{E} \vdot \dd{\vb{A}}$\\
			Komar Two-Form ($K_{\m\n}$) & Field Strength Tensor ($F_{\m\n}$) \\
			Cosmological Constant ($\Lambda$) & Background Charge ($\rho_{\text{bg}}$) \\
			Komar Shift & Background Subtraction\\
			Smarr Formula & Generalized Gauss's Law \\
		\end{tabular}
	\end{ruledtabular}
\end{table}

Furthermore, this exact mapping provides a rigorous resolution to the volume and boundary divergences associated with asymptotically anti-de Sitter spacetimes. As demonstrated in Sec. \ref{sec:cosm}, rather than a naive cancellation of infinite integrals, these divergences are systematically cured by applying Gauss's law to a regularized single-copy field. This gauge-theoretic background subtraction perfectly mirrors the modification of the Komar two-form via a Killing potential. Crucially, the thermodynamic volume term ($PV$) required by black hole chemistry emerges naturally as the flux of this subtracted background field evaluated at the event horizon. 

Moreover, as demonstrated in Sec.~\ref{sec:bardeen} for the Bardeen spacetime, this framework seamlessly accommodates non-linear matter sectors. Even when the matter Lagrangian explicitly depends on global parameters and alters the first law, the resulting non-trivial bulk energy term $\Delta$ is faithfully reproduced by the single-copy charge distribution, providing a transparent electrostatic interpretation of the geometric Komar volume integral.

This foundational setup opens several immediate avenues for future investigation. First, it is crucial to study the effects of non-minimal matter couplings. When such couplings produce a black hole with secondary hair, there is no additional global charge associated with it. It remains an open question how this specific thermodynamic structure is reflected in the single-copy flux calculations. Conversely, if the non-minimal coupling yields a black hole with primary hair, the standard Smarr formula must be explicitly modified. Observing how this modification emerges from the gauge theory source terms will be a highly illuminating exercise.

Second, lower-dimensional gravity provides a unique and necessary testing ground for our framework. In three dimensions, the introduction of a negative cosmological constant is a strict requirement for the existence of black hole solutions (e.g., the Ba\~{n}ados-Teitelboim-Zanelli black hole \cite{Banados:1992wn}). Furthermore, the thermodynamic structure is highly distinctive; in the Smarr formula, either the mass term or the $PV$-term is missing. Moreover, obtaining the Coulomb electric field in the single-copy gauge theory inherently requires matter coupling \cite{CarrilloGonzalez:2019gof,Gumus:2020hbb,Alkac:2021seh,Alkac:2022tvc}. Investigating how these intricate features manifest within our flux-to-thermodynamics dictionary and accurately reproduce the three-dimensional Smarr relation is an important outstanding problem.

Finally, the most significant extension of this work is the generalization to stationary, rotating spacetimes, such as the Kerr and Kerr-Newman black holes. While the introduction of angular momentum breaks spherical symmetry and introduces additional complications, this thermodynamic structure is expected to generalize. Consequently, the rotational work terms (e.g., $2\Omega_H J$) should naturally emerge from the magnetic-type fluxes of the rotating single-copy gauge field. We hope to report on these issues soon elsewhere.

\begin{acknowledgments}
	The author would like to express his sincere gratitude to his colleagues and co-authors for their invaluable collaboration and insights during the research and publication of previous works on the classical double copy and black hole thermodynamics. Their contributions have been fundamental to the development of the ideas presented in this work.
	
	Additionally, the author acknowledges the use of Gemini 3 Flash for assistance in editing and improving the writing of the manuscript, particularly in the Introduction and Conclusion sections. All scientific content, mathematical proofs, and final interpretations remain the sole responsibility of the author.
\end{acknowledgments}
\bibliography{ref}
\end{document}